\documentclass[apj]{emulateapj}
\usepackage{float}
\usepackage{epsfig}
\usepackage{graphicx}
\usepackage{graphics}
\usepackage[latin1]{inputenc}
\usepackage{latexsym}
\usepackage{footnote}
\usepackage[bottom]{footmisc}
\usepackage{amssymb, amsmath}
\newcommand{\nd}{{\nodata}}

\begin{document}
\submitted{Accepted for Publication in the Astronomical Journal}
\title{A Wide-Field Survey for Transiting Hot Jupiters\\ and Eclipsing Pre-Main-Sequence Binaries \\ in Young Stellar Associations\altaffilmark{1}}

\author{Ryan J.~Oelkers\altaffilmark{2,3*}, Lucas M.~Macri\altaffilmark{2}, Jennifer L.~Marshall\altaffilmark{2,4}, Darren L.~DePoy\altaffilmark{2,4},\\ Diego G.~Lambas\altaffilmark{5,6}, Carlos Colazo\altaffilmark{5}, Katelyn Stringer\altaffilmark{2,7}}

\altaffiltext{1}{This paper includes data taken at The McDonald Observatory of The University of Texas at Austin.}
\altaffiltext{2}{George P.~and Cynthia W.~Mitchell Institute for Fundamental Physics and Astronomy, Department of Physics and Astronomy, Texas A\&M University, College Station, TX 77843, USA}
\altaffiltext{3}{Current Address: Department of Physics and Astronomy, Vanderbilt University, Nashville, TN 37235, USA}
\altaffiltext{4}{Charles R.~and Judith G.~Munnerlyn Astronomical Laboratory, Texas A\&M University}
\altaffiltext{5}{Observatorio Astron{\'o}mico, Universidad Nacional de C{\'o}rdoba}
\altaffiltext{6}{Instituto de Astronom{\'i}a Te{\'o}rica y Experimental, IATE-CONICET}
\altaffiltext{7}{Department of Physics and Astronomy, Middle Tennessee State University}
\altaffiltext{*}{Corresponding author, {\tt ryan.j.oelkers@vanderbilt.edu}}

\begin{abstract}
The past two decades have seen a significant advancement in the detection, classification and understanding of exoplanets and binaries. This is due, in large part, to the increase in use of small-aperture telescopes ($<20$~cm) to survey large areas of the sky to milli-mag precision with rapid cadence. The vast majority of the planetary and binary systems studied to date consist of main-sequence or evolved objects, leading to a dearth of knowledge of properties at early times ($<$50~Myr). Only a dozen binaries and one candidate transiting Hot Jupiter are known among pre-main sequence objects, yet these are the systems that can provide the best constraints on stellar formation and planetary migration models. The deficiency in the number of well-characterized systems is driven by the inherent and aperiodic variability found in pre-main-sequence objects, which can mask and mimic eclipse signals. Hence, a dramatic increase in the number of young systems with high-quality observations is highly desirable to guide further theoretical developments. We have recently completed a photometric survey of 3 nearby ($<$150~pc) and young ($<$50~Myr) moving groups with a small aperture telescope. While our survey reached the requisite photometric precision, the temporal coverage was insufficient to detect Hot Jupiters. Nevertheless, we discovered 346 pre-main-sequence binary candidates, including 74 high-priority objects for further study.
\end{abstract}

\section{Introduction}

The first planet to be detected using the transit method \citep{Charbonneau2000, Henry2000} sparked an interest to use the technique on a massive survey scale. Over the past seven years, {\it Kepler} and {\it CoRoT} have dominated the field of transiting exoplanet detection, with {\it Kepler} alone yielding 962 confirmed planets and nearly another 3786 candidates \citep[Exoplanet Orbit Database,][]{Baglin2006, Borucki2010}. The rush to detect and categorize numerous planet candidates has led to the discovery of many unexpected objects whose origins have yet to be fully explained. Of particular interest are ``Hot Jupiters'' (hereafter, HJs), which were among the first extra-solar planets (hereafter, exoplanets) to be detected \citep{Mayor1995}. Their large mass (close to that of Jupiter, $\sim 300 M_\oplus$) and short orbital periods (P$<10$~d) present a challenge to the extant theories of planet formation, many of which were based on the properties of our own solar system. The leading hypothesis to explain the existence of HJs is that they formed beyond the ``snow line'' of their proto-planetary disks ($\ge 4$~AU for a Sun-like star) before migrating inwards \citep{Ida2008}. However, the timescale for this process has yet to be determined observationally or fully explained theoretically. 

Few wide-field, exoplanet surveys have focused on distinct stellar groups or regions for specific science objectives. The second/current phase of the {\it Kepler} mission (called {\it K2}) has a wide variety of science goals including the study of young stars and exoplanets, AGN variability, asteroseismology and supernovae \citep{Howell2014}. In contrast, most ongoing ground-based surveys \citep[such as HAT, KELT and WASP;][]{Bakos2002, Pepper2007,Pollacco2006} target individual bright stars over the entire sky. A dedicated search of young stellar associations is necessary to fully chronicle the formation and migration of HJs.

\begin{deluxetable*}{lccccc}[hbp!]
\tabletypesize{\tiny}
\tablewidth{0pt}
\tablecaption{Observation Log\label{tb:obs}}  
\tablehead{\colhead{Young Stellar} & \multicolumn{2}{c}{Center of the Master Frame} & \colhead{Number of}    & \colhead{Number of} & \colhead{Number of stars}\\ 
                  \colhead{Association}    &\colhead{R.A. [hrs]}           & \colhead{Dec. [deg]}          & \colhead{useful hours} & \colhead{baseline days} & \colhead{in the master frame}}
\startdata
USco & 16:05:42.4 & -24:25:59  & 97.8 & 435 & 104845 \\ 
IC\,2391 &  08:38:37.6 & -53:16:31 & 36.7 & 169 & 108964 \\ 
$\eta$~Cha & 08:43:48.0 & -79:02:07 & 75.6 & 320 & 81046
\enddata
\end{deluxetable*} 

An excellent understanding of circumstellar disk formation, accretion and dissipation is critical to establish the timescales for planet formation and migration. Yet, the failure to discover authentic HJs around pre-main-sequence stars (hereafter, PMS) is in sharp contrast with the expectations from migration models \citep{Yu2015}. The expectation from these models is that planet migration would occur in $<10$~Myr given the time it takes for a typical planetary disk to form, accrete and dissipate \citep{Mamajek2009}. While some theorists have suggested an \textit{in-situ} formation for these objects, they have yet to fully explain how they could retain their primary atmosphere in the warmer environment in front of the snow line \citep{Ida2008,Batygin2015}. Planet scattering is also supported by the wide range of misaligned stellar-spin orbit angles observed among known HJs \citep{Anderson2016}. In this scenario, a large Jupiter-sized planet forms with a similar mass outer companion. These two objects experience a Kozai scattering and the smaller body is ejected while the larger body migrates towards the host star. Currently, only one T-Tauri star ($\sim3$~Myr) is known which possibly hosts a HJ \citep{vanEyken2012}, but this claim is hotly debated \citep{Ciardi2015, Kamiaka2015, Yu2015}. Therefore, the most reasonable test of planetary migration timescales would be a significant increase in the number of detected young HJ candidates.

Searches for young, transiting HJs would also be particularly sensitive to the detection of pre-main-sequence eclipsing binaries (hereafter, PMBs). Precise and accurate measurements of stellar masses and radii at diverse ages, obtained via double-lined eclipsing binaries, provide the most rigorous tests of stellar evolution models \citep{Torres2010, Baraffe2015}. Presently, the vast majority of the systems that have been properly characterized contain main-sequence or evolved objects. In contrast, only a dozen PMBs have been discovered and studied in depth \citep{MoralesCalderon2012}.

Studies of these few PMBs have shown significant differences with predictions, calling into question some of the assumptions adopted by the models. For example, the transformation of observed properties such as temperature and luminosity into mass and age has been shown to be discrepant by $50-100$\% for stars below 1M$_{\odot}$ an inconsistency which can only be relaxed by the use of empirical relations \citep{Stassun2014}. At a broader scale, the determination of star formation rates is very sensitive to the assumed initial mass function -- a parameter that is heavily dependent on the adopted evolutionary tracks of pre-main sequence stars and currently fails to explain the observed distribution of stellar masses \citep{KennicuttEvans2012}. Hence, a significant increase in the number of well-characterized young systems spanning the widest possible range of masses and ages is the best approach to test and eventually improve evolutionary models. 

In this publication, we present initial results from a search for exoplanets and PMS binaries in three nearby young stellar associations using a wide-field, small-aperture telescope. \S2 describes the instrument, targets and observations; \S3 details the data reduction steps; \S4 explains the techniques used to search for variability, periodicity and eclipses; \S5 presents our results and \S6 contains our conclusions.

\section{Survey Details}

\begin{figure*}[tp!]
\centering
\includegraphics[scale = 0.45]{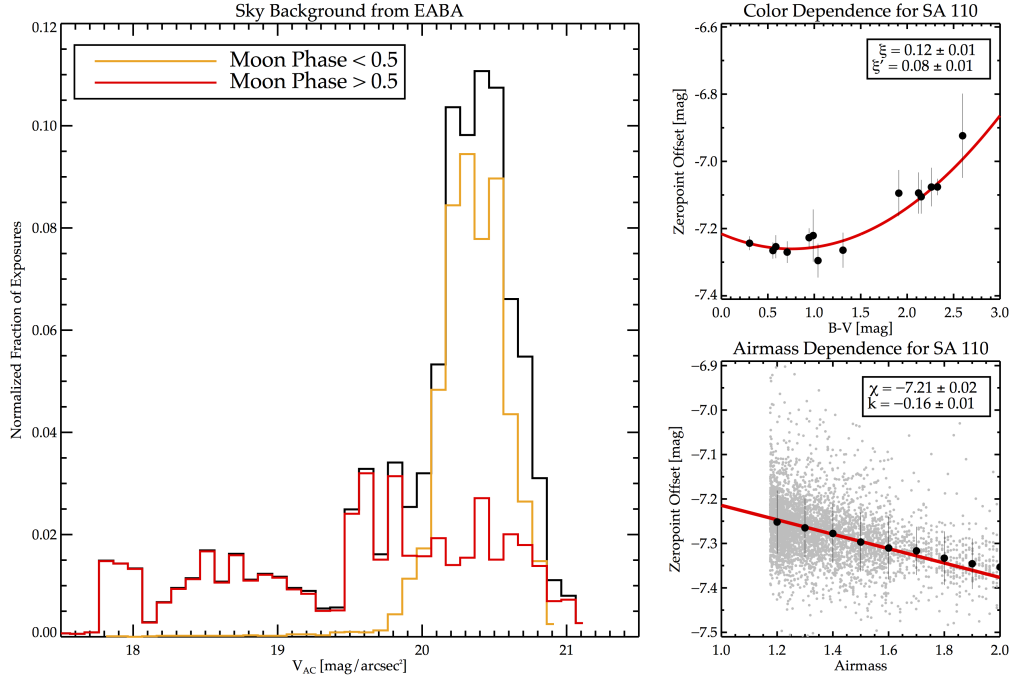}
\caption{\textit{Left}: The sky background from EABA during the survey in mag/arcsec$^2$. The site has generally dark observing conditions especially when the moon phase is $<0.5$. \textit{Top Right}: The zero-point offset of the \textit{AggieCam} detector as a function of B-V color. \textit{Bottom Right} The zero-point offset of the \textit{AggieCam} detector as a function of airmass. All data was calibrated to Johnson-Bessell V band \citep{Johnson1953, Bessell1990} using the Landolt standard star field SA~110 \citep{Landolt1992} using $M_{V_{\rm AC}}=m_{inst}+2.5log_{10}(t_{exp})+\chi+k(X-1)+\xi(B-V)+\xi'(B-V)^2$.\label{fig:eaba}} 
\end{figure*}

\subsection{Instrument}
Our survey instrument, nicknamed \textit{AggieCam}, consists of an Apogee Alta F16M camera with a $4096\times4096$ pixel Kodak KAD-16083 CCD that is thermoelectrically cooled down to $\delta$T$=-45^{\circ}$~C relative to ambient. Testing of the CCD showed a dark current of 0.2 e$^-$/pix/s at temperatures of $-25^{\circ}$~C relative to ambient. The optics include a Mamiya photographic 300~mm lens with a Hoya UV and IR cut filter to restrict the wavelength range to 0.4-0.7$\mu$m. The effective aperture size of the telescope is 53.6~mm and the total throughput of the system is near 45\%. The pixel scale of the detector is $6.2\arcsec$/pix, leading to a total field of view (hereafter, FoV) of $\sim50$~sq.~deg. The telescope was installed at the Estaci{\'o}n Astrofis{\'i}ca de Bosque Alegre (hereafter, EABA) as part of an ongoing collaboration with the Universidad Nacional de C{\'o}rdoba which owns and operates the site.

EABA is a research and outreach observatory located at $31.412^{\circ}$~S, $64.489^{\circ}$~W at an altitude of 1350~m, $\sim50$~km from the city of C{\'o}rdoba, province of C{\'o}rdoba, Argentina. Nearly all observations were carried out remotely from the Mitchell Institute of Fundamental Physics and Astronomy at Texas A\&M University in College Station, Texas. Logistical support for the instrument was provided by staff members of the Instituto de Astronom\'\i a Te\' orica y Experimental, Observatorio de C\' ordoba, and EABA.

\subsection{Targets and Observations}

We targeted three young stellar associations to maximize the science return from our study: IC\,2391 ($\alpha=8^{\rm h} 40^{\rm m}, \delta=-53^\circ$), the $\eta$~Chamaeleontis cluster ($\eta$~Cha, $\alpha=8^{\rm h} 45^{\rm m}, \delta=-79^\circ$), and the Upper Scorpius association (USco, $\alpha=16^{\rm h}, \delta=-24.5^\circ$). Given the spread in R.A. between IC\,2391 and USco and the circumpolar nature of $\eta$~Cha, we were able to make an efficient use of the camera all year long.

The USco association is a subsection of the larger Scorpius-Centarus association and it is the nearest OB association to the Sun \citep{Preibisch2002}. At a distance of 145~pc and a mean age of $11\pm1\pm2$~Myr \citep{Pecaut2012} it is an excellent candidate for our survey. This association was chosen specifically for its large O/B/A membership ($\sim$100 confirmed members) which implies thousands of unconfirmed low-mass members \citep{Rizzuto2015}. 

IC\,2391 is a large, loose, young open cluster in Vela. Measurements of the main-sequence turn off and the lithium depletion boundary suggest an age of 30-50 Myr, while {\it Hipparcos} and Tycho-2 data led \citep{Reipurth2008} to determine the distance to this cluster to be $147\pm5$~pc. This association was chosen because of its older population; we may expect to see transiting HJs if migration timescales are constrained by disk dissipation. 

Lastly, $\eta$~Cha is a cluster located at a distance of 97~pc with confirmed stellar ages between $2-18$~Myr \citep{Mamajek1999}. $\eta$~Cha provides a variety of stellar ages to probe planetary formation, has a high number of confirmed T-Tauri stars and was chosen to ensure we could observe at least one target all year. 

Individual exposures were set to 60~s and were taken when a field was above an airmass of 3. Table~\ref{tb:obs} denotes the number of exposures taken and baseline of days between the first and last exposure for each field. For the remainder of this paper we will use the abbreviated names of each target (USco, $\eta$~Cha \& IC\,2391) to refer to all stars in the respective fields of view and not just the {\it bona-fide} members.

\section{Data Analysis}

\begin{deluxetable*}{lccccccrrc}
\tabletypesize{\tiny}
\tablewidth{0pt}
\tablecaption{\textit{AggieCam} Stellar Library* \label{tb:lib}}  
\tablehead{\colhead{Star ID} & \multicolumn{2}{c}{Coordinates J2000} & \colhead{V$_{AC}$}    & \multicolumn{3}{c}{Metrics} & \multicolumn{2}{c}{Period [d]} &  \colhead{Cluster Type}\\
                      &\colhead{R.A. [hrs]}           & \colhead{Dec. [deg]}          & & V & P & E & \colhead{LS} & \colhead{BLS} & }
\startdata
U098597&16:03:30.09&-24:31:47&11.043&1&1&0&  2.530870&  \nd &STU\\
U038698&16:12:38.88&-23:26:56&14.804&1&1&0&  1.618511&  \nd  &IRV\\
V027962&08:39:10.08&-55:47:08&13.040&1&1&0&129.932922&  \nd  &LTN\\
V046511&08:40:56.08&-55:09:34&13.663&1&1&0&  0.594557&   \nd &IRV\\
U048565&16:11:19.20&-26:29:15&15.174&1&1&0&  1.354051&  \nd  &IRV\\
U089470&16:04:58.48&-26:14:12&15.433&1&1&0&  3.987664&  \nd  &STU\\
V038184&08:35:21.32&-55:27:32&13.831&1&1&0&  1.011477&  \nd  &STU\\
U050223&16:10:44.16&-23:03:57&14.983&1&1&0&255.915588& \nd   &LTU\\
U096707&16:03:52.96&-26:48:23&13.949&1&1&0&271.910309&  \nd  &LTU\\
V060026&08:51:16.05&-54:38:32&12.686&1&1&0& 21.655487& \nd   &IRV\\
U089370&16:05:01.97&-26:58:09&14.720&1&1&0&  1.022940&   \nd &STU\\
C068057&09:25:32.43&-78:34:45&13.512&1&1&0&114.386337&  \nd  &LTN\\
U046582&16:11:26.73&-24:28:02&10.436&1&1&0& 24.036270& \nd   &STU\\
U057495&16:09:42.50&-24:31:03&13.074&1&1&0&  6.360475&   \nd &STU\\
V028652&08:34:04.63&-55:46:34&12.986&1&1&0& 45.440865&  \nd  &LTN\\
U024137&16:15:19.63&-24:18:42&12.614&1&1&0&255.915588&  \nd  &LTU\\
V038671&08:29:06.09&-55:26:53&12.741&1&1&0& 39.282047& \nd   &STU\\
C015455&08:35:06.59&-81:31:09&11.980&1&1&0&103.316689&  \nd  &LTN\\
V079454&08:38:14.68&-54:07:23&13.961&1&1&0&  0.174279&   \nd &STU\\
U088883&16:05:06.32&-26:55:37&16.064&1&1&0& 12.220644& \nd   &STU\\
V081626&08:43:48.46&-54:01:37&13.038&1&1&0&  0.156274&  \nd  &STU\\
... & ... & ... & ... & ... & ... &... & ... & ... & ... 
\enddata
\tablecomments{*:The full table is available for download with the online publication. A 1 in the metric column denotes if the star passed the variability (V), periodicity (P) or eclipse (E) testing in \S~4. The types in the Cluster Type column are based on the 5 clusters described in \S~\ref{subsubsec:class}; STU is short-term uniform periodic variability, IRV is infrequent variability, STN is short-term non-uniform periodic variability, LTU is long-term uniform variability, LTN is long-term non-uniform variability. Additionally, TRN is an expired transit candidate.}
\end{deluxetable*} 

\subsection{Pre-Processing}

Each frame was pre-processed for bias subtraction, flat fielding and residual background correction. The master bias frame was generated by median-combining 160 bias frames taken throughout the observing season. The master flat field was created by median-combining 410 sky flats taken throughout the observing season with values $\ge 10000$~ADU. We bias-subtracted, scaled and median-combined the selected frames to make a temporary flat field, applied it to the images, masked any stars and repeated the process to generate the final flat. This master flat frame was normalized by the center $2048\times2048$ pixels to avoid contamination by the vignetted corners of the detector. 

We applied a residual background correction following the approach of \citet{Wang2013}, \citet{Oelkers2015} to remove any variations due to clouds, moonlight or scattered light. The residual background model is constructed by sampling the sky background every $64\times64$ pixels over the entire detector. Bad or saturated pixels are excluded from each sky sample. A model sky is then fit inside each box and interpolated between all boxes to make a thin plate spline \citep{Duchon1976}. We used the IDL implementation GRID$\_$TPS to make the spline which is subtracted from the frame. 

We corrected for slight drifts in tracking by carrying out aperture photometry on all images using {\tt DAOPHOT} \citep{Stetson1987}, matching the resulting star lists with {\tt DAOMATCH}, and solving for the geometric transformations with {\tt DAOMASTER}. We aligned the images using the bi-cubic interpolation implemented in the IDL routine {\tt POLY$\_$WARP}.

\subsection{Difference Image Analysis}

We used Difference Imaging Analysis \citep[hereafter, DIA;][]{AlardLupton} to match the point-spread function (PSF) of each science frame and a reference master image and measure changes in stellar flux. Since the PSF in our images was spatially varying, we applied a $5\times5$~pix, $2^{\rm nd}$-order, Dirac-$\delta$-function kernel across the frames \citep{Alard2000,Miller2008}. A detailed description of the algorithm is given in \citet{Oelkers2015}, and the code is publicly available from https://github.com/ryanoelkers/DIA. Stamps were taken around bright, isolated stars to solve for the kernel coefficients using the least-squares method.

\subsection{Flux Extraction}

We extracted the stellar flux on each differenced frame using the IDL implementation of the {\tt DAOPHOT} routine {\tt APER}. The aperture radius was set to 3~pix (18.6$\arcsec$) with a sky annulus from 5-7~pix (31- 43.4$\arcsec$). These fluxes were added to the flux from the master frame, zero-pointed and corrected for exposure time. We only selected stars located between 500 \& 3500 in both (x,y) in the master frame to remove objects with poorer photometry due to vignetting. The final star lists are given in Table~\ref{tb:lib}.

We used ensemble photometry to identify and remove systematic trends due to instrumental or processing effects that were present in multiple light curves. We used the Trend Fitting Algorithm \citep{Kovacs2005} as implemented in the {\tt VARTOOLS} package \citep{Hartman2008} to compensate for these systematics. We used 100-250 stars spanning a wide range of fluxes and (x,y) positions that did not exhibit any discernible variations of an astrophysical nature (e.g., eclipsing binaries, periodic variables, etc.) as templates for the trend removal. 

\subsection{Photometric calibration}
We determined the transformation of \textit{AggieCam} magnitudes, obtained through a fairly wide filter, to the standard $V$ band via observations of SA~110 \citep{Landolt1992} at airmass values of $1.18 < X < 2.0$. The field was selected because of the relative large number of bright standard stars in our FoV (14) which spanned a large range in color ($0.3 < B\!-\!V < 2.6$). Fluxes were extracted following the same procedure used for our science observations. We solved for the following transformation equation:

\begin{equation*}
	m_{{\rm V}_{\rm {AC}}}=m_{\rm inst}+2.5log_{10}(t_{\rm exp})
\end{equation*}
\begin{equation}
	+\chi+k(X-1)+\xi(B-V)+\xi'(B-V)^2
\end{equation}
where $m_{{\rm V}_{\rm {AC}}}$ is the {\it AggieCam} $V$-band calibrated magnitude; $m_{\rm inst}$ is the instrumental magnitude, $t_{exp}$ is the exposure time of the observation; $\chi$ is the zero-point offset; $k$ is the airmass coefficient; $X$ is the airmass of the observation; $\xi$ and $\xi'$ are the first- and second-order color terms, respectively; and $(B-V)$ is the color of the star in the standard $B$ and $V$ bands. We solved the transformation equation using the IDL routines {\tt POLY$\_$FIT} and {\tt LINFIT}, finding $\chi=-7.21\pm0.02$, $k=-0.16\pm0.01$, $\xi=0.12\pm0.01$ and $\xi'=0.08\pm0.01$ as shown in Figure~\ref{fig:eaba}.

We determined the sky background at the observatory by calculating the median pixel value of each image. We find the background is fairly low for an observatory in relatively close proximity to a major metropolitan area. The lowest values were found towards the South, where the instrument was primarily pointed, when the lunar phase was $<0.5$. The median sky background was $V=20.5$~mag/sq. arcsec, comparable to the median value at Mauna Kea (20.7 in the same units). Figure~\ref{fig:eaba} shows the normalized number of observations and their median sky background values.

\subsection{Noise}

\begin{figure*}[tp!]
\centering
\includegraphics[scale=0.5]{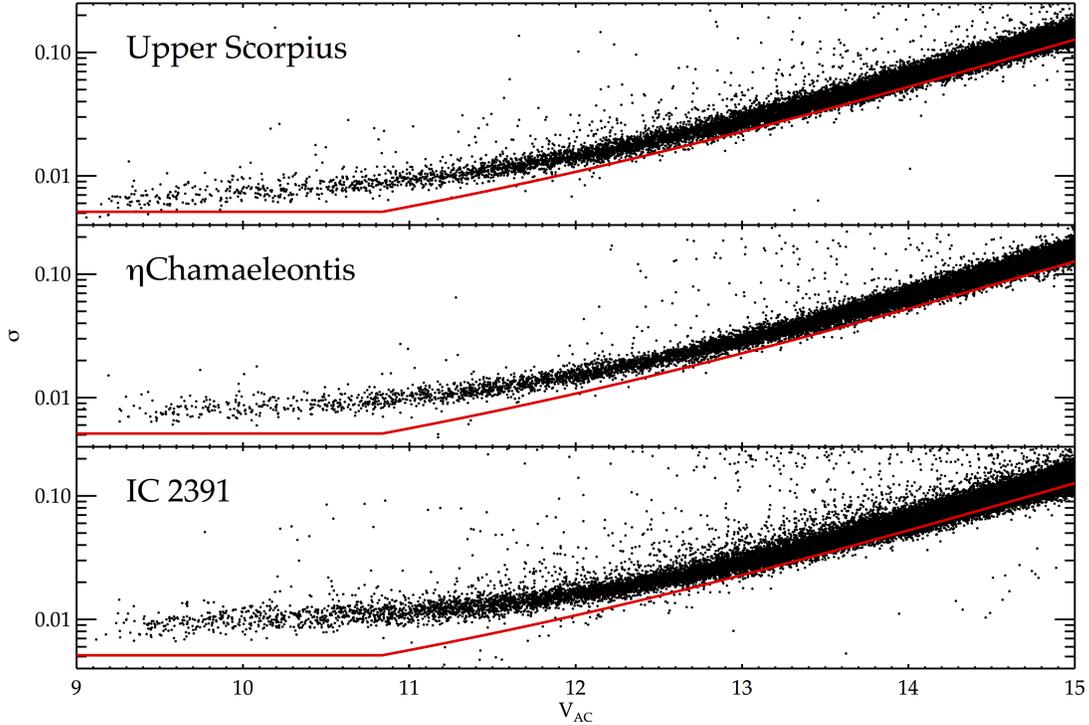}
\caption{The achieved photometric precision of \textit{AggieCam} from a typical day of observations. The red line denotes the expected noise from the star, sky, detector and scintillation limit.\label{fig:disp}}
\end{figure*}

One of the leading reasons PMS stars have been so poorly studied is due to their intrinsic stellar variability, which appears to be both erratic and aperiodic. This variability is likely the result of ongoing stellar contraction towards its main-sequence radius, increased spot cycle due to magnetic activity, proto-stellar disk accretion or a combination of all these factors \citep{Stassun2014}. This variability can mask and mimic signals which are important to the understanding of the fundamental properties of stars such as determining stellar rotation rates, identifying planetary \& stellar eclipses or making precise radial velocity measurements.

Studies attempting to reduce the impact of astrophysical variation in PMS stars rely on the blind whitening of light curves against suspected periodic variability \citep{Kraus2015} or on the use of self-described, overly-flexible data-driven models which de-weight variation similar to the desired signal \citep{WangD2015}. While these techniques adequately remove variation, the possibility that \textit{bona-fide} signals will be removed is greatly increased. Therefore, it is necessary to isolate and understand the source of each signal and the associated noise, prior to removal, in order to preserve the scientific integrity of the data.

We modeled the statistical uncertainty as $\sigma^2=I_{N}+A I_{sky}+2A(RN^2)+\sigma^2_a$, where $I_{N}$ and $I_{\rm sky}$ are the photon counts from the object and sky respectively, $A$ is the area of the photometric aperture, RN is the read noise of the detector and $\sigma_a$ is the expected scintillation limit defined by \citet{Young1967}, \citet{Hartman2005} as:

\begin{equation}
S = S_0 d^{-\frac{2}{3}}X^{\frac{7}{4}}e^{-h/8000}(2t_{\rm ex})^{-\frac{1}{2}}
\label{eq:scin}
\end{equation}

\noindent{where $S_0 \sim 0.1$~mag, $d$ is the telescope diameter in cm, $X$ is the airmass, $h$ is the altitude in m and $t_{\rm ex}$ is the exposure time in s. The values for {\it AggieCam} at EABA are: $d=5.36$, $h=1350$, $1<X<3$ and $t_{\rm ex}=60$. We find $0.003 < S < 0.017$~mag, with a value of 5~mmag for the median airmass of our observations ($\langle X\rangle=1.5$).

We measured the dispersion in each light curve, weighted by the uncertainty in aperture photometry, for a single night of observations (typically $60-360$ frames) and compared these values to our noise model. We found satisfactory agreement with the simple model described above, with dispersions of $1-1.4\times$ the scintillation limit for stars with V$_{\rm AC}<10.9$ as shown in Figure~\ref{fig:disp}.

\section{Searching for Variability, Periodicity and Eclipses\label{sec:tests}}

\begin{figure*}[tp]
\centering
\includegraphics[scale=0.474]{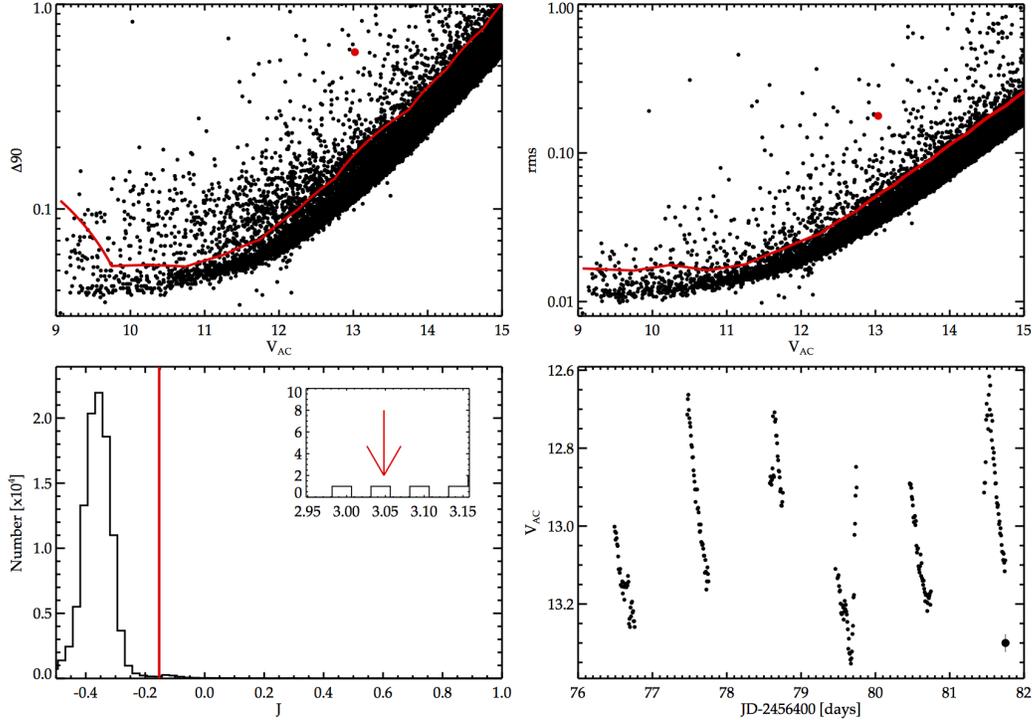}
\caption{The variability tests used to identify variable candidates in each young stellar association. Stars lying above the red line in the top panels and to the right of the line in the bottom left panel are expected to be variable. \textit{Top Left}: $\Delta_{90}$ statistic with the upper 2$\sigma$ quartile plotted as a red line. \textit{Top Right}: rms statistic with the upper 2$\sigma$ quartile plotted as a red line. \textit{Bottom Left}: J Stetson statistic with the upper 3$\sigma$ cut plotted as a red line. \textit{Bottom Right}: The light curve of the variable candidate U071728 from the USco field. The candidate is shown clearly passing each statistic as a red dot in the top two panels and a red arrow in the bottom left panel. The light curve is shown in 10~m bins with the size of the typical photometric error shown at the bottom right.\label{fig:stat}}
\end{figure*}

\begin{figure*}[bp]
\centering
\includegraphics[scale=0.474]{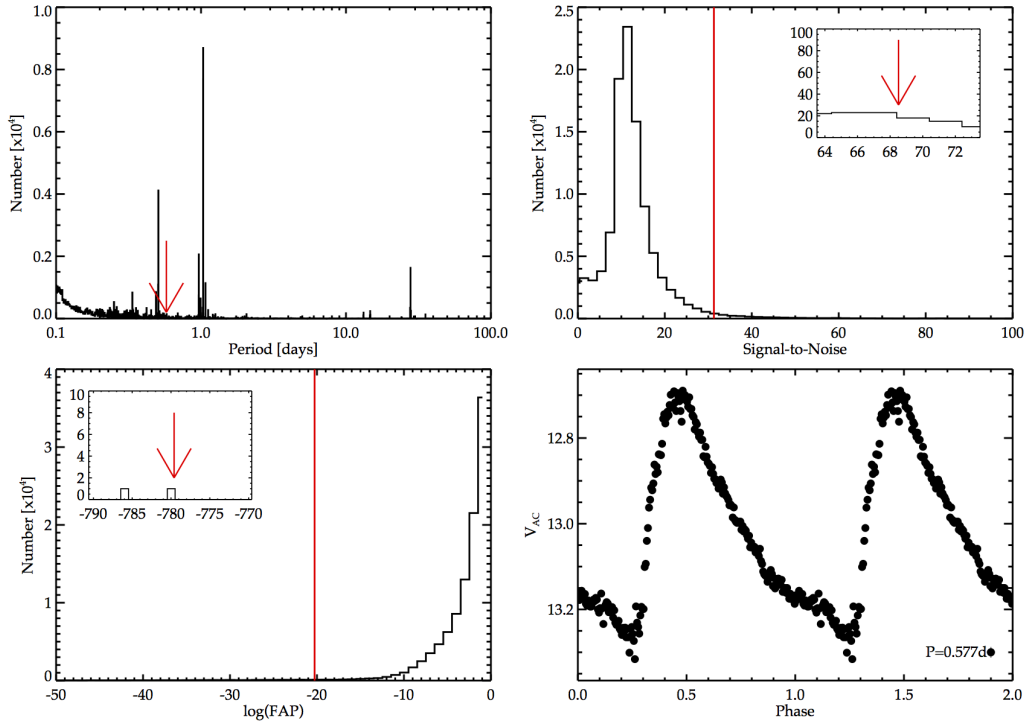}
\caption{The periodicity tests used to identify periodic candidates in each young stellar association. \textit{Top Left}: the number of ``periodic'' stars with similar periods, indicative of aliasing. The passing candidate is shown with a red arrow. Notice the period is not found on or near a large distribution of other periods; \textit{Top Right}: the $+3\sigma$ cut (red line) on the signal-to-noise ratio. The passing candidate is shown with an arrow in the window insert; \textit{Bottom Left}: the $-3\sigma$ cut (red line) on the false alarm probability. The passing candidate's log$_{10}$(FAP) is shown with an arrow in the window insert; \textit{Bottom Right}: The light curve of a periodic variable star candidate U071728 (also shown in Fig~\ref{fig:stat}). The light curve has been phase folded on the recovered period of 0.577~d, binned into 200 data points and plotted twice for clarity. The typical error is shown at the bottom right of the panel. \label{fig:perd}}
\end{figure*}

\subsection{Classical Variability and Periodicity\label{subsec:tests1}}

We employed a combination of 3 variability metrics, following the approach of \citet{Wang2013}, \citet{Oelkers2015}. First, we computed the root-mean-square (hereafter, rms) of all stars and the upper 2$\sigma$ envelope as a function of magnitude; objects lying above this limit are likely to be genuine astrophysical variables. Next, we computed the magnitude range spanned by 90\% of the data points of every light curve (hereafter, $\Delta_{90}$) and its upper 2$\sigma$ envelope as a function of magnitude. Since we wished that both statistics be based on ``constant'' stars only and not be biased by large-amplitude variables, both envelopes were calculated in an iterative fashion. 

Finally, we computed the Welch-Stetson $J$ variability statistic \citep[hereafter, $J$;][]{Stetson1996} including the necessary rescaling of DAOPHOT errors \citep{Kaluzny1998}. $J$ is useful to detect variability during short time spans, such as those sampled by the 60~s cadence of \textit{AggieCam}, since it computes the significance of photometric variability between two or three adjacent data points. $J$ is expected to produce a distribution of values with a mean value close to zero for the ``constant'' stars and a one-sided tail towards positive values for the ``variable'' stars. We considered objects lying above the $+3\sigma$ value as variable.

We also applied a variety of cuts to our variable sample to ensure we were not contaminated by stars showing high dispersion due to systematics. We removed the 10\% of stars with lowest number of data points and we required each star to be farther than 5 pixels of a star 2 magnitudes or brighter.  We \textit{only} consider a star to be variable if it passes all 3 of these metrics. Figure~\ref{fig:stat} shows these techniques recovering the variable candidate U071728. 

We searched each light curve for periodic signals using a Lomb-Scargle periodogram \citep[hereafter, LS;][]{Lomb, Scargle} as implemented in {\tt VARTOOLS} \citep{Hartman2008, Press1992, Zechmeister2009}. We computed the highest SNR period of each star between 0.1~d and the total number of baseline days of observation for each association. We removed periods within 0.001~d of any of the first 10 harmonics (\textit{f, f/2, f/3, ...}) of the sidereal day (0.99726958~d) to alleviate contamination from the most common observing alias. We also removed periods which were within 1~d of the lunar sidereal month, 27.32~d. We applied $+3\sigma$ cuts on the false alarm probability, $\log_{10}({\rm FAP})$, and cumulative SNR. The FAP provides an estimate on the likelihood of a true periodic signal by comparing the SNR of a specific signal to the cumulative distribution of all SNRs for each peak in the LS periodogram. Figure~\ref{fig:perd} shows these techniques recovering the periodic variable candidate U071728.

\subsection{Stellar \& Planetary Eclipses\label{subsec:tests2}}

We ran the Box Least Squares algorithm \citep[hereafter, BLS;][]{Kovacs2002} to search for eclipse-like events of the desired HJs. The signal is non-sinusoidal because the transit is only expected to occur for a very short amount of phase, typically $<10$~\% \citep{Charbonneau2000}. The BLS routine searches for signals caused by a periodic alternation between two flux levels. The probability of detecting a small, periodic, eclipse-like feature is greatly increased by iteratively searching over the parameter space of period, eclipse depth and transit length. 

We also ran a BLS search, where prior to the search, we pre-whitened each light curve against the primary LS period between 0.1 and 10~d and its 10(9) (sub-)harmonics. We searched each light curve for transit candidates with a primary BLS period between 0.1 and 10~d with a transit length of 0.01 and 0.1 of the primary BLS period. We chose the limit of 10~d because HJs, by definition, do not have periods longer than 10~d. We allowed for $10^4$ trial periods and 200 phase bins. We also adopted a number of detection thresholds that are common among exoplanet searches. We required no less than 3 transit events for every candidate to ensure no significant variation between the odd and even eclipses, which would suggest an eclipsing binary over a planetary transit. While typical planetary transits produce a drop in the light curve of only 1 to 2\%, we kept larger depth events since they could be due to other interesting objects such as brown dwarfs or eclipsing binaries.

We then subjected each light curve to criteria based on the statistics of the BLS routine. Typically, the error in ground-based milli-magnitude photometry is correlated. Because this is the regime in which we searched for exoplanets, we investigated the signal-detection-efficiency statistic (hereafter, SDE) described in \citet{Kovacs2002} to determine the significance level of each transit. Any transit candidate with a SDE statistic greater than 4 was considered significant. True transits should only show the systematic dimming of each light curve and not a systematic brightening or anti-transit. \citet{Burke2006} suggests a transit to anti-transit statistic $\Delta \chi^2 / \Delta \chi_{-}^2$, where both the transit and anti transit $\chi^2$ values are compared. Stars with the statistic $\geq1.0$ were considered candidates. Finally, all candidates passing these statistics were visually inspected to confirm the eclipse-like variation. If the star was known to be PMS from a previous study \citep{Rizzuto2015} then we allowed the star to fail one of the two statistics because its \textit{a priori} membership made it a high priority candidate. We refined each period by searching the first 5 harmonics of the BLS period incase the routine was triggering on the wrong frequency.

\subsection{Higher-Precision Photometric and Preliminary Spectroscopic Followup}

Any transiting HJ or PMB candidate passing all of the significance tests described in \S~\ref{subsec:tests1} \& \S~\ref{subsec:tests2} was then subject to a series of followup photometric observations.  These observations are used to provide independent, higher-precision transit and eclipse measurements. They also helped to confirm and refine the orbital period, depth, ephemeris and duration for each eclipse. The time of secondary eclipse was probed to rule out large variations in flux, which would indicate a grazing EB or background blend. The higher angular resolution of the follow-up telescopes also allowed us to determine any contamination from blending due to the coarser pixel scale (6.2$\arcsec/$pix) of \textit{AggieCam}. These observations also provided color measurements which helped to identify spectral type, binary contamination in the HJ sample and association membership.

The 1.54~m telescope at EABA provided 300+~hrs of {\it BVRI} photometry to date, with further observations planned. The 0.8~m telescope at McDonald Observatory provided 14~hrs of BVRI photometry. The Las Cumbres Global Observatory Telescope Network (LCOGT) provided 30~hrs of \textit{gri} photometry from their 1~m facilities. The Texas A\&M University campus observatory 0.5~m telescope provided 30~hrs of \textit{gri} photometry. Additionally, the 2.1~m telescope at McDonald Observatory, coupled with the Sandiford Echelle Spectrograph \citep[][hereafter, SES]{McCarthy1993}, provided 14~hrs of initial spectroscopic followup.

\section{Results}

\begin{figure*}[tp]
\centering
\includegraphics[width = \textwidth]{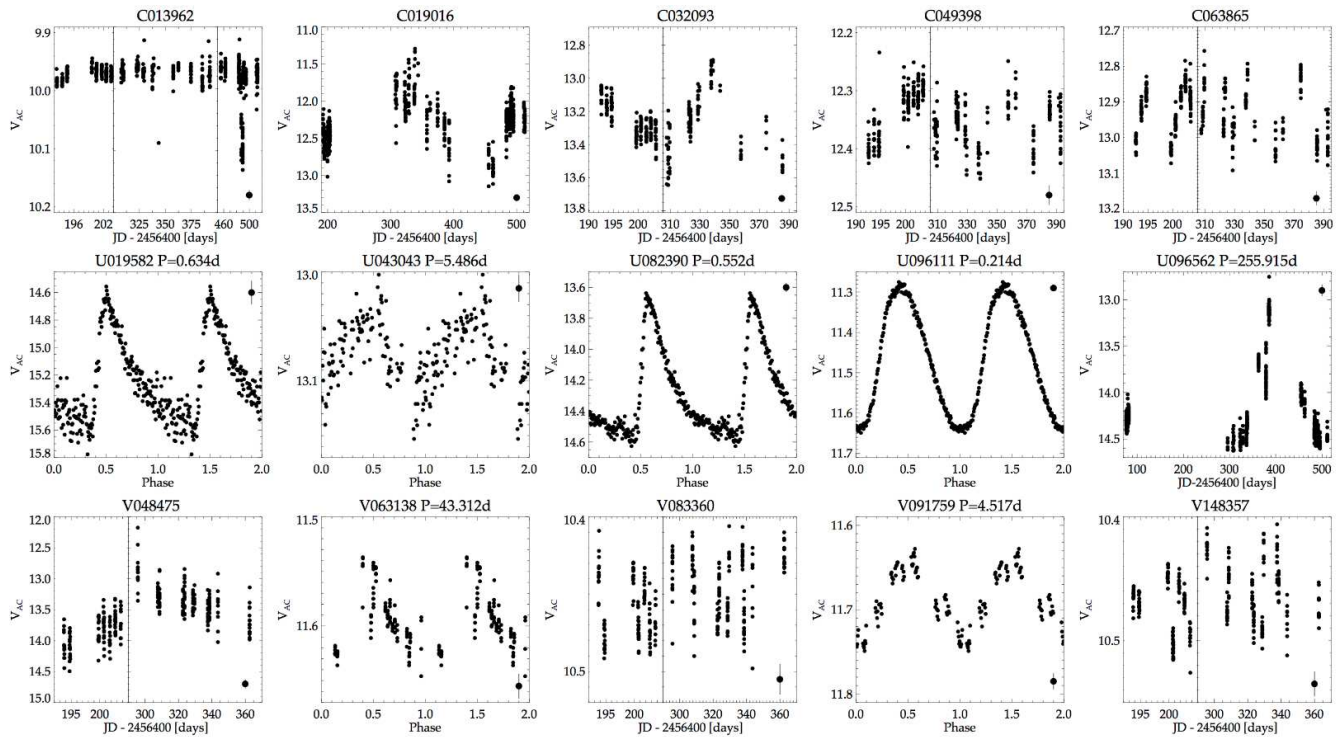}
\caption{15 variables found in each of the associations as part of our search. The period of the star is displayed in the title of the panel if it was identified as significant in our search. Panels in temporal space have been binned in 10~m intervals. Panels in phase space have been binned in 200 phase bins and plotted twice for clarity. Typical photometric error is plotted on the top or bottom right of each panel. \label{fig:vcands}}
\end{figure*}

\begin{deluxetable}{cccc}[bp!]
\tabletypesize{\tiny}
\tablewidth{0pt}
\tablecaption{Distribution of Significant\\ Variable Periods\label{tb:perd}}  
\tablehead{\colhead{Timescale [d]} & \colhead{USco} & \colhead{$\eta$~Cha} & \colhead{IC\,2391} }
\startdata
$<1$          &   851     & 95  &    1062\\
$1-10$        &   337    &  32   &   331\\
$10-100$     &    148  &    41 &     141\\
$>100$         &  112   &   16   &    18
\enddata
\end{deluxetable}

\subsection{Classical Variable and Periodic Stars}

Employing the metrics described above, we identified over 1,599 variable candidates across all 3 fields; 593 in USco, 689 in IC\,2391 and 317 in $\eta$~Cha. We determined a normalized variable star rate of $1.98\pm0.08\times10^{-4}$ ($1.38\pm0.07\times10^{-4}$, $2.22\pm0.08\times10^{-4}$) variable stars per sq.~deg.~in USco ($\eta$~Cha, IC\,2391). These rates are generally consistent with one another and with those  of previous wide-field variable star searches \citep{Wang2011, Wang2013, Oelkers2015, Oelkers2016}.

Similarly, we found 3,184 stars showing significant periodicity using the cuts above. We found 240 (62; 127) stars with significant variability that also exhibited a periodic signal in USco ($\eta$~Cha; IC\,2391). Table~\ref{tb:perd} presents the number of stars in each field with periodic variations of different timescales. Figure~\ref{fig:vcands} shows 15 candidate variable and periodic stars from all 3 fields. 

\subsubsection{Classification of Variable Stars\label{subsubsec:class}}

We classified each variable star with a general type using an IDL implementation of a k-means clustering algorithm, {\tt CLUST$\_$WTS} and {\tt CLUSTER}. k-means clustering attempts to segregate features of each object into k-clusters to minimize the distance between each point in the cluster and the center of the cluster \citep{Macqueen1967}. We selected 6 features of each variable star light curve to use in the clustering algorithm:

\begin{enumerate}
 \item The magnitude difference between the mean magnitude of the first half of the light curve and the total mean magnitude.
 \item The magnitude difference between the mean magnitude of the second half of the light curve and the total mean magnitude.
 \item $J$, described in \S~\ref{subsec:tests1}.
 \item The skewness of the light curve.
 \item The kurtosis of the light curve.
 \item The LS period, if it was determined to be significant as described in \S~\ref{subsec:tests2}.
\end{enumerate}

Features 1 \& 2 were selected to aid in identifying variable stars showing an overall increasing, decreasing or ``constant'' trend with time. Feature 3 was selected since its measure of variability is insensitive to  light curve dispersion, as opposed to rms or $\Delta_{90}$. Features 4 \& 5 were selected to describe the uniformity of the change in magnitude around the mean magnitude (i.e. large constant oscillations or infrequent pulses). Finally, feature 6 was selected to aid in differentiating between long- or short-period variations. 

The variable stars were clustered into 1 of 5 types using the features above. We found selecting more clusters made too many indistinguishable clusters and selecting fewer clusters grouped too many diverse objects.
 
 \begin{enumerate}
 \item Short-term, uniform periodic motion (i.e RR~Lyrae, contact, semi-detached \& short-period detached binaries, sinusoidal variables).
 \item Infrequent variability (i.e possible flares, long-period detached binaries, possible systematics).
 \item Short-period, non-sinusoidal periodic variables.
 \item Long-term, sinusoidal periodic variation.
 \item Long-term, irregular variation.
\end{enumerate}
 
Each variable star has been identified with its cluster membership in the final stellar library described in Table~\ref{tb:lib}. We note that many stars in cluster 2 appear to show variation at identical times, likely due to systematics of the detector. We have elected to include them in our sample because they passed all of the variability cuts outlined in \S~4. We visually classified stars in clusters 1 \& 2 to identify the eclipsing binary candidates described in \S~\ref{subsec:binmem}.
\begin{figure*}
\centering
\includegraphics[scale = 0.4]{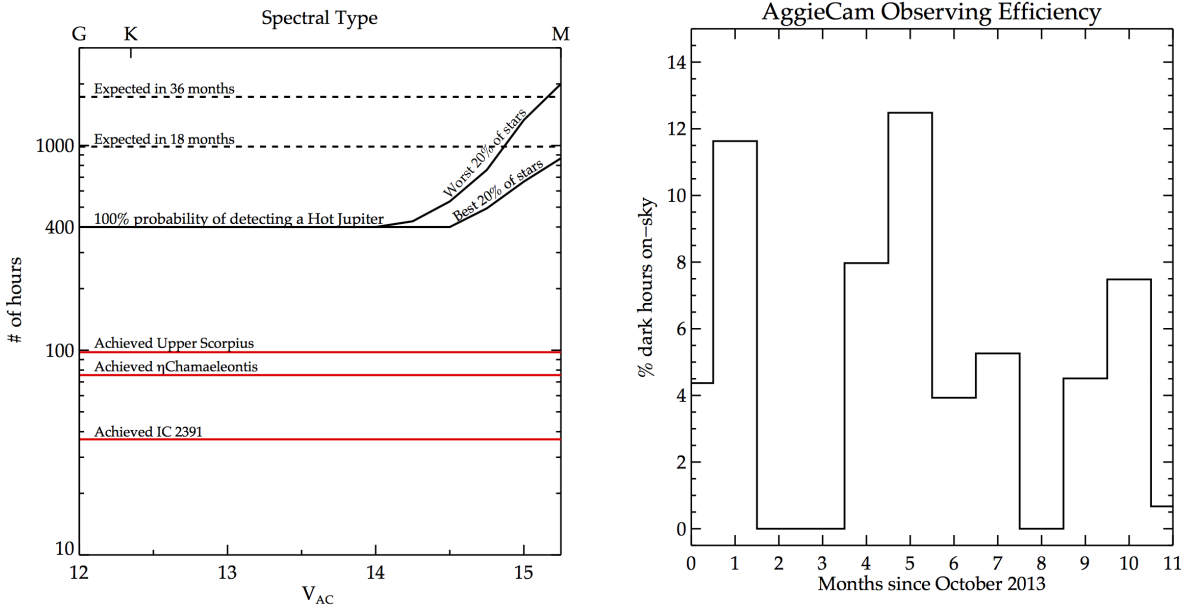}
\caption{\textit{Left}: The black line denotes the cumulative number of hours required to detect a transiting 1.5 $R_J$ HJ with P$=2.7$~d. The split in the black line marks the 20\% of stars with the best and worst rms, respectively. The dotted lines mark our expected cumulative hours of observation prior to the survey. The red lines mark the achieved cumulative observed hours for each association. \textit{Right}: Percentage of dark hours that yielded useful data during each month of the \textit{AggieCam} observing campaign since October 2013. A combination of unprecedented weather conditions, coupled with technical failures, led to a low observing efficiency. \label{fig:eff}}
\end{figure*}

\subsection{Exoplanet Candidates and the Migration Timescale for Hot Jupiters}

\begin{figure*}
\centering
\includegraphics[width=\textwidth]{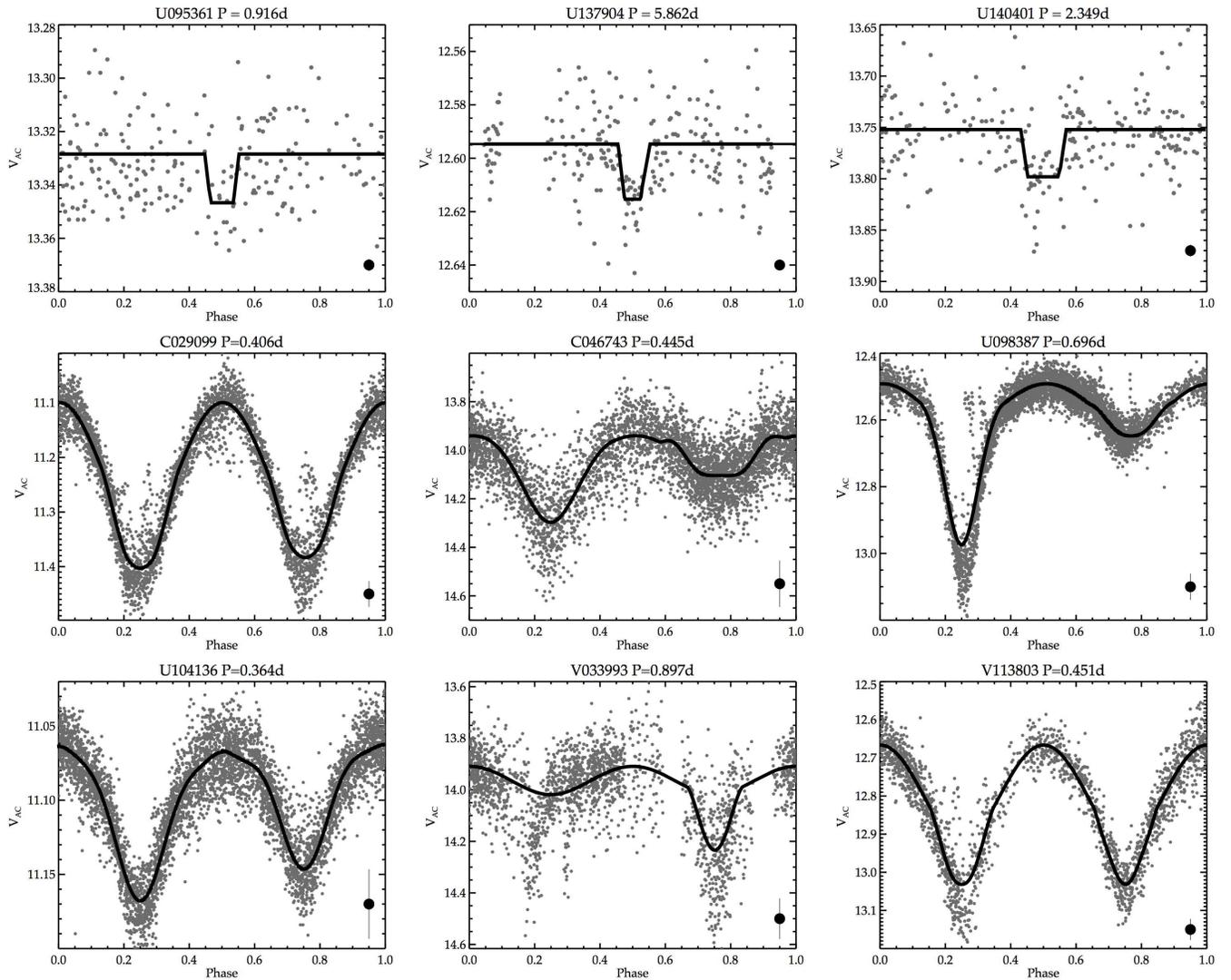}
\caption{[Top Row] Three HJ candidates of various periods, detected at high significance as part of our search. The HJ candidate light curves have been binned into 30~m bins and phase folded. The black line denotes the best-fit BLS model of the transit \citep{Bakos2002}. [Middle \& Bottom Row] 6 PMB candidates from each association. The black lines denote the best fit JKTEBOP model \citep{Southworth2013}. Errors are shown in the bottom right of each figure.\label{fig:cands}}
\end{figure*}

Previous observational studies of main-sequence stars have suggested the detection rate of a transiting HJ to be $\sim 1$ planet for every 5000 stars observed per month \citep{Burke2006}. Our survey was expected to extend far beyond the one-month time span required and we assumed we would detect multiple HJs. While each association does not contain the 5000 stars required to find one transiting HJ in one month's time, the planned survey duration of 3 years and knowledge of cluster members would allow selective targeting to determine the presence of young transiting HJs. It was also expected we would find transiting HJs around stars that have not yet been confirmed as cluster members due to the dearth of effective wide-field surveys of each association.

As a proof of concept, we injected transit signals for a variety of HJs with a range of periods ($0.5<P<10$~d) and radii ($0.8<R_J<2.0$) into simulated light curves with dispersions based on the typical rms limits of the detector and calculated the number of hours of observation required for a detection that passed the criteria outlined above. We used the published distance to each association and the 30 Myr isochrones from \citep{Pietrinferni2006} and calculated the stellar mass and radius for a range of apparent magnitudes. Figure~\ref{fig:eff} shows the results of this calculation for a 1.5~$R_J$ with a period of 2.7~d. Taking into account long-term weather statistics for the site, we expected that after 18 months of observations we would be able to collect $\sim 400$~hrs of data and achieve the required sensitivity over most of the target stellar types (F to early K). A further 18 months of observations would extend our sensitivity to the $K/M$ boundary.

The combination of an unprecedented rainy season at EABA and equipment and server failures severely reduced our on-sky efficiency, and the survey was prematurely terminated after 15 months due to a lightning strike. Only $\approx 200$ hours of observations were obtained. Our monthly efficiency rate is shown in Figure~\ref{fig:eff}, along with the final survey numbers for each association ($\sim100, 75, 40$~hrs for USco, $\eta$~Cha and IC\,2391). 

Even with our low number of observations, we identified 7 possible planetary eclipse events as shown in Figure~\ref{fig:cands}. We performed higher-precision photometric followup of each planetary candidate with LCOGT and EABA to confirm each eclipse and the expected ephemeris of mid-transit. We also used archive photometry from the \textit{K2} Campaign 2 mission \citep{Howell2014}. Unfortunately, we were unable to recover the eclipse events at the expected ephemeris times. We have included a list of each of the candidates and their determined orbital information for completeness in Table~\ref{tb:hjs}.

We need to properly assess our ability to detect these objects and identify any potential biases, before we can speculate on the scientific implications of our null result. We will make this interpretation for USco \textit{only}, because of the significantly larger number of candidates recovered (6) and the larger number of observed hours compared to $\eta$~Cha and IC\,2391. We will do this by answering two questions to determine the confidence on our null result: (1) How many HJs should exist in this associations? (2) If a HJ exists, would we have detected it?

\subsubsection{How many Hot Jupiters should exist in this association?}

\begin{deluxetable*}{lccccccc}
\tabletypesize{\tiny}
\tablewidth{0pt}
\tablecaption{Hot Jupiter Candidates \label{tb:hjs}}  
\tablehead{\colhead{Star ID} &  \multicolumn{2}{c}{Coordinates in the Master Frame} & \colhead{BLS Period}    & \colhead{BLS Ephemeris} & \colhead{Mag.} &\colhead{Depth}                  & \colhead{Follow-Up} \\ 
                       			    &\colhead{R.A. [hrs]}           & \colhead{Dec. [deg]} & \colhead{[days]}            & \colhead{JD-24546400}    & \colhead{V$_{\rm AC}$} & \colhead{mmag} & \colhead{Observatory}}
\startdata
U022979	& 16:15:39.5 & -24:55:41 & 0.818758 & 76.631834 & 13.407  & 27 & Bosque Alegre\\ 
U064813	& 16:08:35.9 & -23:55:39 & 0.706294 & 77.273527 & 12.737  & 28 & LCOGT \\ 
U095361	& 16:04:06.7 & -26:37:07 & 0.915534 & 77.040402 & 13.331 & 16 & LCOGT\\
U120649	& 16:00:07.3 & -26:16:46 & 1.493101 & 76.612173 & 14.197  & 28 & Bosque Alegre\\
U137904	& 15:57:16.7 & -25:29:19 & 5.861766 & 80.164762 & 12.597 & 17 & K2 \\
U140401	& 15:56:55.5 & -22:58:40 & 2.349168 & 77.197127 & 13.754 & 42 & Bosque Alegre\\
C067591  & 07:50:41.6 & -78:43:34 & 0.251447 & 192.535710 & 13.463 & 29 & Bosque Alegre
\enddata
\end{deluxetable*} 

\begin{figure}[h]
\centering
\includegraphics[scale = 0.33]{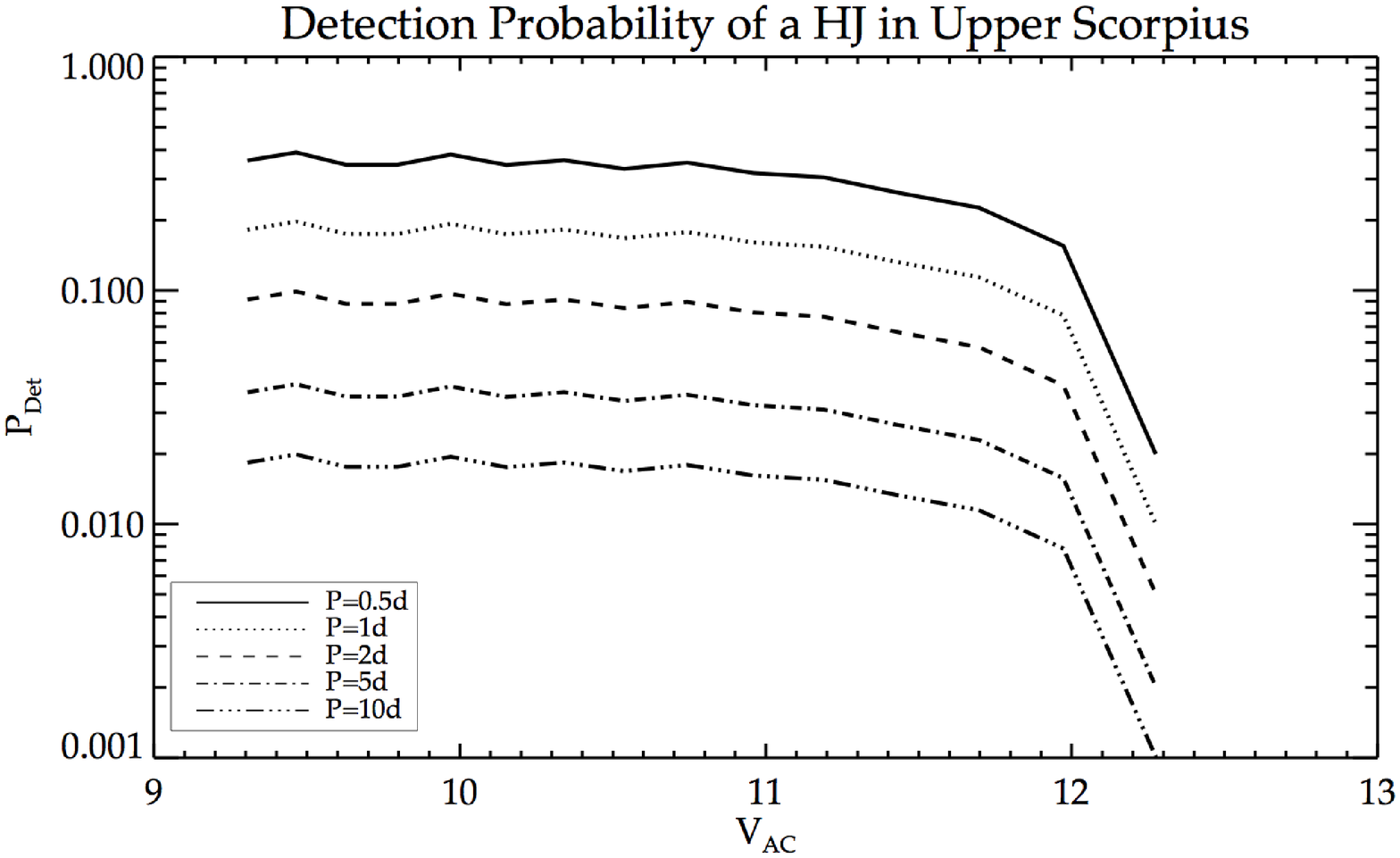}
\caption{The probability of detecting a HJ in our sample as a function of magnitude for stars in USco. The probability of detecting a HJ with P$<0.5$~d is $\sim34\%$ for $V_{\rm AC}<12$ and quickly declines for longer periods and fainter magnitudes. \label{fig:probdet}}
\end{figure}

\begin{figure}[h]
\centering
\includegraphics[scale = 0.33]{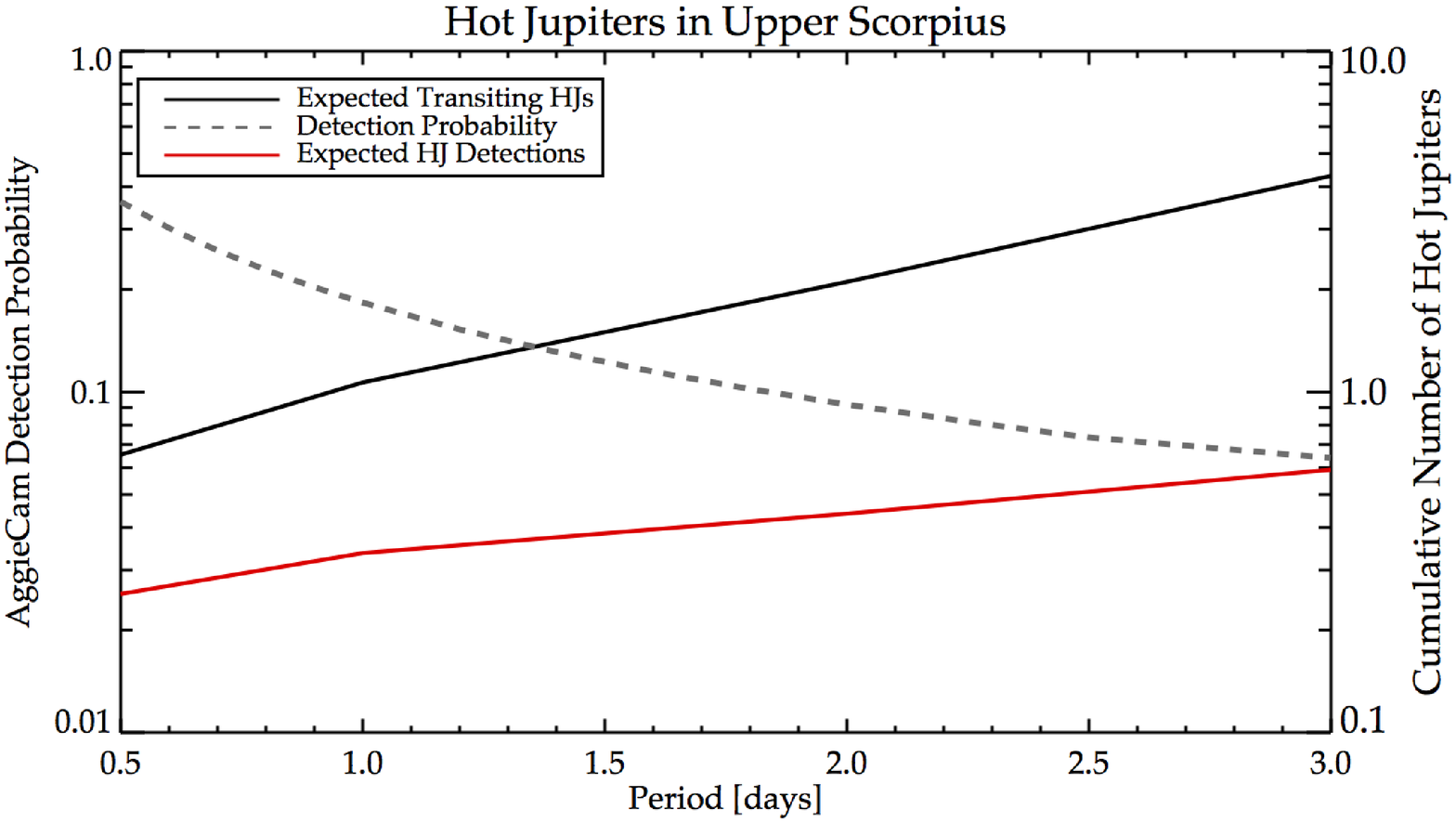}
\caption{The cumulative number of expected HJs in the USco association as a function of period. The black line denotes the number of HJs detected with 100$\%$ efficiency while the red line denotes our expected detection rate. The grey dashed line is our detection efficiency as a function of orbital period. At $P=0.5$~d where, we are most likely to detect planets, we find our null result is consistent with expectation that $<1$~HJ would exist with such a period. \label{fig:numdet}}
\end{figure}
Previous studies of main-sequence stars have determined the planet fraction of HJs is $P_{\rm frac}\approx 1\%$. However, the expectations for Very Hot Jupiters (P$<3$~d) are $P_{\rm frac}\approx 0.1\%$ \citep{Hartman2008, Bayliss2011, Fressin2013}. Given the target associations have 100-1000 confirmed members, extra care needs to be given to interpreting our null result. By combining these planet fractions with the detection probabilities below, we can determine if our null result is consistent with the expectation of HJ migration due to disk interactions.

\subsubsection{If a Hot Jupiter exists, would we have detected it?}

We determine our detection probability following the same logic as \citet{Burke2006}, with the full detection probability being defined as

\begin{equation}
	P_{\rm det,i} =  \int \int \frac{d^2p}{dR_pda}P_{\epsilon, i}(a,R_p)P_{T,i}(a,R_p)P_{\rm mem,i}dR_pda
\end{equation}
 
\noindent where $R_p$ is defined as the planet radius; \textit{a} is the semi-major axis; $P_{\epsilon, i}(a,R_p)$ is the probability of detecting a transit around the $i^{\rm th}$ star in the survey averaged over orbital phase and inclination; $P_{T,i}(a,R_p)$ is the probability a planet of $R_p$ and $a$ would transit its host star; $P_{\rm mem,i}$ is the probability the $i^{\rm th}$ star is a member of the association. 

We calculate $P_{\epsilon, i}$ following the logic of \citet{Gaudi2000} and integrate over the survey's achieved photometric precision, achieved observing cadence and the minimum threshold to detect HJs of various periods ($0.5<P<10$~d) and radii ($0.8<R_J<2$). The calculation of $P_{T,i}$ is independent of our survey data. If we assume a random inclination orientation of a star with respect to the observer then $P_{T,i} = \frac{R_{*}+R_p}{a}$ where $R_{*}$ is the radius of the star. We assumed a uniform distribution of inclination angles for our calculation. We integrated these detection limits over all stars in our FoV because we were also interested in detecting planets around non-association members, if they existed.

\subsubsection{Implications of the Null Result}

If we have not detected any HJs because they have not yet migrated, and its not simply an artifact of the low number of observations, we can attempt to understand the implications for the migration timescale. Because disk formation, accretion and dissipation is expected to be finished by 10~Myr, it is surprising the planets have not yet migrated to their `Hot' positions. Both Type I and II migration require an actively accreting disk to move the planet inwards towards its host star \citep{Ida2008}. Similarly, recent studies have also suggested \textit{in-situ} HJ formation and that the snow line may not be a requirement for primary atmosphere retention \citep{Batygin2015}. 
 
Recent work by \citet{Mann2016} has shown the detection of a $\sim5$~R$_\oplus$ `Hot Neptune' with an orbital period of $\sim$5.4~d orbiting an 11~Myr, M3 star in USco. While the star is not in our FoV, it is unlikely we would have detected the eclipse given the parameters of the system. The star's apparent magnitude is $15.5$ in \textit{V} and the planet transits with a $\sim0.003$~mag eclipse depth (we transformed APASS \textit{g}\&\textit{r} to V using the relations from \citet{Lupton2005}; APASS V for this star was determined to be unreliable \citep{Henden2012}). \textit{AggieCam's} precision at this magnitude is $>0.1$~mag (see Figure~\ref{fig:disp}). We also have a very low recovery rate for periods greater than 1.0~d (see Figure~\ref{fig:probdet}) and our null detection at these periods matches the expectation for these objects in this field (see Figure~\ref{fig:numdet}). Similarly, \citet{JohnsKrull2016} have identified a HJ candidate (P$\sim$9~d) through optical and infrared radial velocity measurements of the PMS star CI Tau (2~Myr). These detections provide evidence that disk migration is the driving force for gas-giant migration at early times ($<11$~Myr) and helps to confirm the observations of previously discovered young, transiting, `hot', gas-giants \citep{vanEyken2012}. These detections also imply we did not detect a HJ only because we failed to reach our desired temporal coverage of the field and not because these objects have failed to migrate before 11~Myr.

\subsection{Binary Candidates and Membership Confidence Testing\label{subsec:binmem}}

Our survey yielded 346 PMB candidates; 151 in USco, 138 in IC\,2391 and 57 in $\eta$~Cha. All of these objects require higher-precision photometric and spectroscopic followup for firm classification and accurate determination of stellar parameters. Since the spectroscopic followup of more than 300 objects is not a trivial task, we ranked the candidates by priority on a scale of 0 (low) to 7 (high). We calculated the ranking using archival 2MASS \& WISE colors to determine infrared excesses \citep{Skrutskie2006, Wright2010}, using the UCAC4 proper motion catalogue to compare the motion of a given candidate with the mean of its putative association \citep{Zacharias2013}, inferring possible component masses based on eclipse depth \& apparent magnitude and high-resolution spectroscopy to search for the Li I line at 6708~\AA~for 7 initial test binaries. The final results of these tests are shown in Table~\ref{tb:bins}.

\subsubsection{Color Selections}

An infrared excess is used to describe an object which appears to have a normal spectral energy distribution in the ultraviolet and visible wavelengths but shows a large excess of flux at infrared wavelengths. These excesses manifest themselves in young stellar objects because these stars are typically enshrouded in a dusty proto-stellar disk which absorbs ultraviolet light and re-emits it in the infrared. 

To determine which stars may show these infrared excess, we investigate the 2MASS and WISE archival photometry \citep{Skrutskie2006, Wright2010}. We combined the $J\!-\!H$ vs.~$H\!-\!K$ color-color diagram with the stellar locus provided by \citet{Pecaut2013} to estimate the spectral type of each binary. If the star was within 2$\sigma$ of a particular point in the locus we deemed the photometry to be accurate enough to estimate the spectral type. 

Following the methodology of \citet{Luhman2012} \& \citet{Rizzuto2015} we made color cuts using photometry in the 2MASS $K_s$ and the {\it WISE} W2, W3 and W4 bands. Stars lying above the lines denoted in Figure~\ref{fig:excess} are flagged as showing infrared excesses indicative of a younger stellar population. These boundaries are calculated based on the expected photospheric flux observable \textit{after} dissipation of the proto-planetary disk. Binaries showing these excesses were flagged as high-priority candidates, with each excess adding an additional point to the total priority score. We found 11 (10, 21) stars in USco ($\eta$~Cha, IC\,2391) to have at least 1 infrared excess. 

\subsubsection{Proper Motions}

We used the proper motions from the UCAC4 catalogue \citep{Zacharias2013} to identify objects which show similar motions to the well-studied O/B/A/F stars in each association. Using the measured mean proper motions of each association from previous studies \citep{Dodd2004, Luhman2012, Marti2013} and the expected $15-20$~mas error of the UCAC4 catalogue, we identified all candidate binaries which were within 3$\sigma$ of the mean proper motion. Only stars with measured proper motions were included in our significance testing. We increased the priority score for each candidate based on the difference, in $\sigma$, between the star's proper motion and the mean of the association as follows: $\Delta\mu<1\sigma=+3$; $1\sigma<\Delta\mu<2\sigma=+2$; $2\sigma<\Delta\mu<3\sigma=+1$; $\Delta\mu>3\sigma=0$. If the candidate had no measured proper motion, it was given no points. We found 45 (10, 55) stars in USco ($\eta$~Cha, IC\,2391) to have a proper motion within 3$\sigma$ of the proper motion of the cluster. Figure~\ref{fig:promo} shows measured proper motions for the binary candidates in $\eta$~Cha and their positions in the \textit{AggieCam} FoV. 

\subsubsection{Testing Viable Binary Components}

\begin{figure}[h]
\centering
\includegraphics[scale = 0.2]{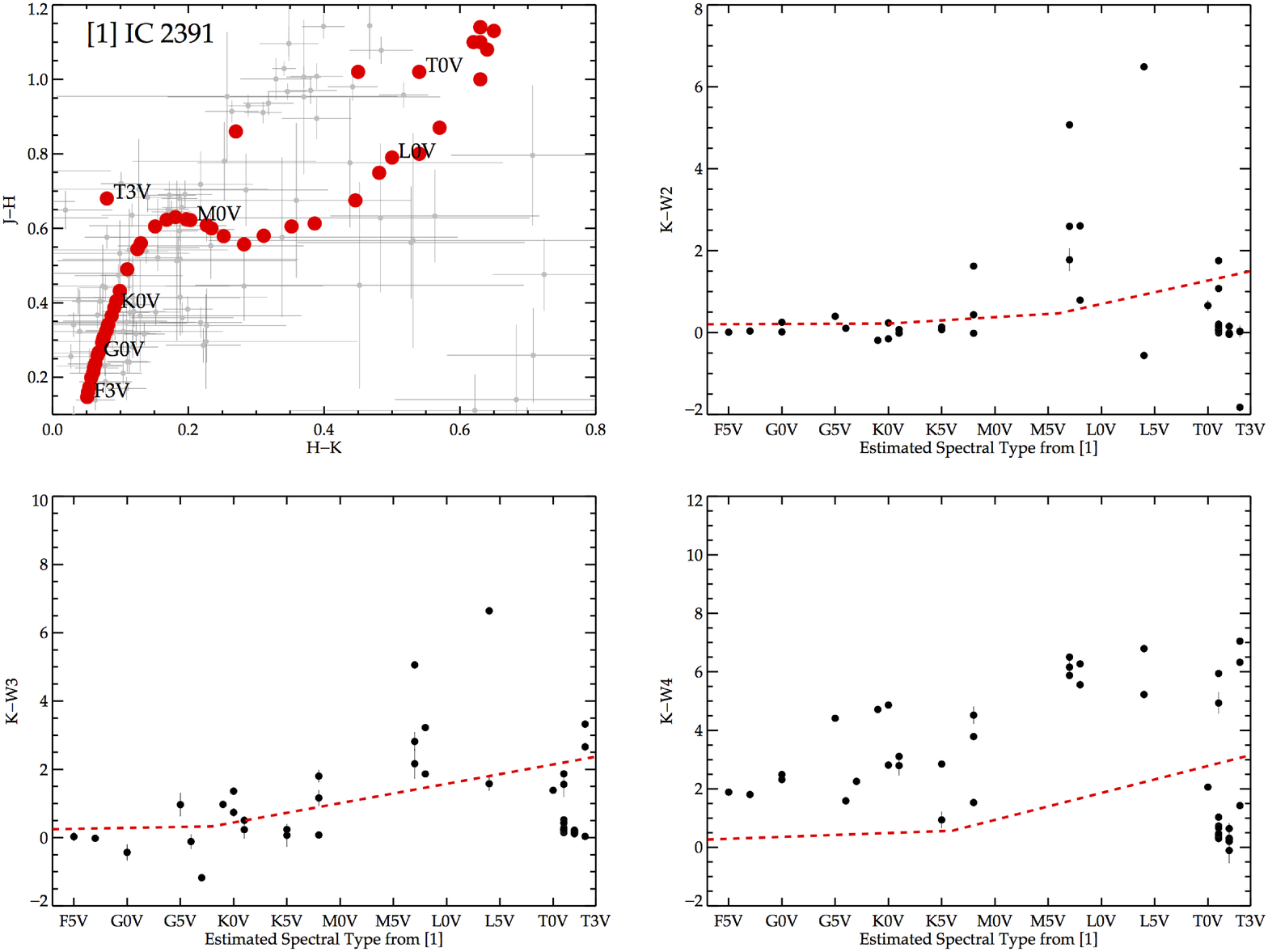}
\caption{The infrared excesses of IC\,2391. [Top Left] 2MASS color-color diagram and stellar locus from \citet{Pecaut2013}. Stars were assigned with a specific spectral type if they were located within 2$\sigma$ of a particular data point in the locus. [Top Right, Bottom Left \& Bottom Right] (K-W2 (W3,W4)) vs estimated spectral type diagram. Stars lying above the red lines are considered to have an excess amount of flux in the infrared, indicative of an accretion disk. The red dotted line marks where the expected photospheric excess ends for main-sequence objects \citep{Luhman2012, Rizzuto2015}. \label{fig:excess}}
\end{figure}

\begin{figure}[h]
\centering
\includegraphics[scale = 0.28]{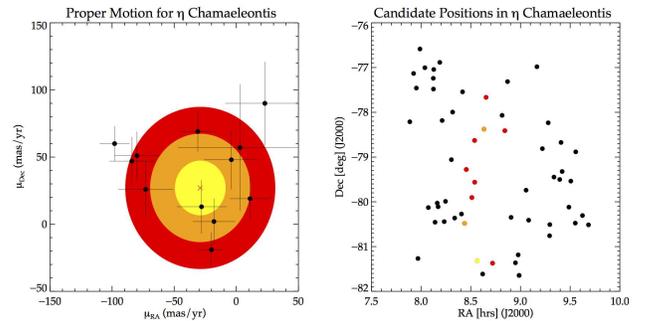}
\caption{[Left] The measured proper motions of stars in the $\eta$~Cha field with measured proper motions from the UCAC4 catalogue plotted with their 1$\sigma$ error bars. Objects were selected as likely candidates based on their position relative to the mean proper motion of the moving group. The colored regions denote 1, 2 \& 3$\sigma$ errors of the UCAC4 catalogue (yellow, orange and red, respectively). [Right] The position of each PMB candidate in the FoV of \textit{AggieCam}. The coloring of the points relates to the proximity of the proper motion of a given star to the mean of the association. A black dot means the star either was too far from the mean motion or did not have a proper motion in UCAC4. \label{fig:promo}}
\end{figure}

\begin{figure}[h]
\centering
\includegraphics[scale = 0.28]{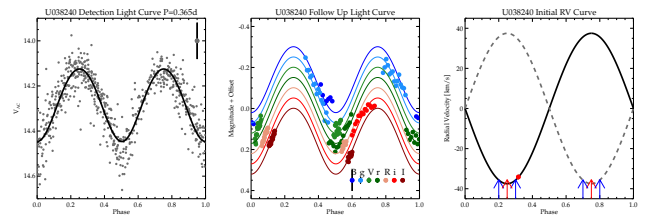}
\caption{PMB candidate U038240 discovered in USco at each stage of the classification process: detection [top]; multi-color followup [middle]; initial RV measurements [bottom]. The solid lines denotes the best-fit JKTEBOP models \citep{Southworth2013}. The expected RV variation is extrapolated from the eccentricity of the system and the best-fit orbital parameters of JKTEBOP. The blue arrows denote the acceptable range of observation in phase and the red arrows denote exact quadrature. \label{fig:binary}}
\end{figure}

\begin{deluxetable*}{lcccc}
\tabletypesize{\tiny}
\tablecaption{Binary Candidate Testing\label{tb:bins}}  
\tablehead{\colhead{Young Stellar} & \colhead{Effective Temperatures \&} & \colhead{Proper Motions}    & \colhead{Infrared Excess}     & \colhead{Total Number}\\ 
                  \colhead{Association}    &\colhead{Separations}                               & \colhead{[UCAC4]}              & \colhead{[2MASS \& WISE]}  & \colhead{of Candidates}}
\startdata
USco 	   & 134 & 45  & 11 & 151 \\ 
IC\,2391   & 131 & 55  & 21 & 138 \\ 
$\eta$~Cha & 47  & 10  & 9  &  57
\enddata
\end{deluxetable*} 

\begin{deluxetable*}{cccccccc}
\tabletypesize{\tiny}
\tablewidth{0pt}
\tablecaption{Pre-Main-Sequence Eclipsing Binary Candidates* \label{tb:pmb}}  
\tablehead{\colhead{\textit{AggieCam} ID} & \multicolumn{2}{c}{Coordinates}                    & \colhead{Magnitude}    & \colhead{Period} & \multicolumn{3}{c}{Metrics} \\ 
                                                         &\colhead{R.A. [hrs]}  & \colhead{Dec. [deg]}  & \colhead{V$_{\rm AC}$}     & \colhead{[days]} & \colhead{ET} & \colhead{IR} & \colhead{PM}}
\startdata
V089825 & 08:39:01.00 & -53:50:25 & 13.442 &  2.025352 & 1 & 3 & 3 \\
V130033 & 08:41:57.79 & -52:31:33 & 12.108 &  0.808794 & 1 & 1 & 3 \\
V065240 & 08:48:14.21 & -54:30:25 & 15.076 &  0.961662 & 1 & 1 & 3 \\
U077182 & 16:06:53.44 & -26:49:19 & 15.019 &  0.535817 & 1 & 2 & 2 \\
V033325 & 08:38:49.10 & -55:36:27 & 13.121 &  0.225700 & 1 & 2 & 2 \\
V096596 & 08:28:07.54 & -53:40:02 & 12.045 &  2.271902 & 1 & 1 & 3 \\
V105161 & 08:54:43.25 & -53:15:18 & 14.289 &  0.821764 & 1 & 2 & 2 \\
V152845 & 08:33:58.26 & -51:32:17 & 11.418 &  5.413995 & 1 & 3 & 0 \\
V167741 & 08:24:08.44 & -50:55:27 & 12.747 & 10.897848 & 1 & 0 & 3 \\
V106482 & 08:23:40.70 & -53:20:25 & 11.745 &  1.804665 & 1 & 3 & 0 \\
V077746 & 08:53:21.85 & -54:03:24 & 13.848 &  1.250303 & 1 & 0 & 3 \\
U104136 & 16:02:38.25 & -24:01:52 & 11.100 &  0.364308 & 1 & 0 & 3 \\
C105839 & 08:02:16.16 & -77:00:14 & 13.275 &  0.881469 & 1 & 3 & 0 \\
U098597 & 16:03:30.09 & -24:31:47 & 11.043 &  5.061739 & 1 & 0 & 3 \\
V105749 & 08:33:29.74 & -53:21:17 & 14.790 &  1.693400 & 1 & 0 & 3 \\
V152442 & 08:26:58.84 & -51:33:55 & 14.116 &  0.211067 & 1 & 3 & 0 \\
V087571 & 08:37:16.21 & -53:53:02 & 13.046 &  1.961866 & 1 & 0 & 3 \\
V142813 & 08:46:07.25 & -51:52:21 & 14.176 &  1.257733 & 1 & 3 & 0 \\
V095317 & 08:27:57.28 & -53:40:32 & 12.327 &  0.437948 & 1 & 3 & 0 \\
U098050 & 16:03:36.56 & -25:09:13 & 12.276 &  1.034986 & 1 & 0 & 3 \\
U056279 & 16:10:04.17 & -26:25:49 & 12.443 &  6.114638 & 1 & 0 & 3 \\
... & ... & ... & ... & ... & ... &... & ... \\
\enddata
\tablecomments{*:The full table is available for download with online publication. The ET metric denotes stars with effective temperature and separations plausible at the distance of the cluster; The IR flag denotes how many IR excesses were identified in the 2MASS \& WISE photometry; PM denotes stars with proper motions consistent with the moving group to 1-3$\sigma$.}
\end{deluxetable*} 

The component masses of the system can still be estimated even without proper spectroscopic observations. The ratio of the eclipse depths is proportional to the fourth-root of the ratio of the temperatures of the photospheres of the stars. This ratio can then be used in conjunction with the apparent magnitude and the distance to the association to place limits on the masses of the components, under the assumption that the object is a {\it bona-fide} member.

We fit a parabola to both the primary and secondary eclipses using the IDL routine {\tt POLY$\_$FIT} to determine the minimum flux value. We then calculated the ratio of the effective temperatures as follows:

\begin{equation}
	({\frac{T_S}{T_P}})^4 =  \frac{F_0-F_P}{F_0-F_S}
\end{equation}

\noindent{where $F_0$ is the flux at quadrature and $F_P$ \& $T_P$ are the flux during the primary eclipse and effective temperature of the primary star respectively (an S denotes the secondary).}

The distance modulus to each association was subtracted from the apparent magnitudes to determine the expected absolute magnitude of each candidate. The \citet{Spada2013} zero-age-main-sequence isochrones were employed to calculate the magnitudes of the components that could be combined to generate this magnitude based on the temperature ratio. Kepler's Third Law was used to calculate the separation of the system using the known period. We determined the viability of each binary based on the orbital separation and the shape of the light curve. For example, a binary composed of two K-dwarfs with a separation of 1 solar radius should be in a nearly contact system and \textit{not} a detached system. If the system was deemed viable it was given 1 point toward its priority score. We found 134 (47, 131) stars in USco ($\eta$~Cha, IC\,2391) to have viable binary components for their expected distance.

\subsubsection{Spectroscopy}

Preliminary spectroscopy was obtained for 7 PMB candidates at McDonald Observatory during the Spring of 2015. The combination of spread in ephemeris timing and poor observing conditions only allowed each PMB to be observed once, close to quadrature. Nevertheless, these measurements provide valuable information about the radial velocity and spectral types for each system. We obtained the spectra using the SES on the 2.1~m at McDonald Observatory. Th-Ar exposures bracketed the science observations and were used to wavelength-calibrate the data using the IRAF routine {\tt REFSPEC}. The radial velocities were calculated using a cross-correlation with reference spectra taken on the same night and spectral templates from SDSS DR5 \citep{Adelman2007}. While the SNR of our spectra were not very high ($<10$), we nevertheless achieved $1-5$km/s precision on the radial velocity measurements. The low SNR prevented any detection of Li~6708~\AA.

We then subjected the detection light curve, followup multi-color photometry and initial estimates of the radial velocity to the JKTEBOP binary-fitting program \citep{Southworth2013}. JKTEBOP models each component of a binary as a sphere for calculating the eclipse shapes and a biaxial ellipsoid for calculating proximity effects. The program uses Levenberg-Marquardt optimization to find the best fit model. An example of a binary being subjected to this fitting process is shown in Figure~\ref{fig:binary}.

\section{Conclusions}

Although our observations were plagued by bad weather, technical setbacks and a lightning strike, we were able to reach the necessary precision for an exoplanet \& eclipsing binary survey. We detected 346 eclipsing pre-main-sequence binary candidates and identified 7 candidate transiting Hot Jupiters; the latter were ruled out with higher-precision followup observations. Additionally, we identified and categorized 4,354 variable and periodic stars across the 3 target associations.

In order to properly interpret our null result, we determined our detection probability to be $34\%$ for HJs with $P<0.5$~d.  Recent work by \citet{Mann2016} has identified a `Hot Neptune' transiting a PMS star in USco, thereby providing evidence for disk migration and suggesting we could have detected transiting planets if we had obtained sufficient temporal coverage of each field.

Finally, we identified 346 candidate pre-main-sequence binaries. 12\% of these stars show an infrared excess, 32\% of these stars have proper motions consistent with their putative association and 90\% have temperature ratios which are consistent with two zero-age-main-sequence stars at the distance of the respective association. We found 74 PMB candidates to have a priority score of 3 or more, denoting a high-likelihood of being pre-main-sequence binaries and have prioritized them for follow-up observations.

\acknowledgements
RJO \& LMM acknowledge financial support from the George P.~and Cynthia Woods Mitchell Institute for Fundamental Physics \& Astronomy and the Mitchell-Heep-Munnerlyn Endowed Career Enhancement Professorship in Physics or Astronomy. They also thank the observing staff at the Estaci{\'o}n Astrofis{\'i}ca de Bosque Alegre and the Observatorio Astronomica de C{\'o}rdoba; specifically Pablo Guzzo, Cecilia Qui{\~n}ones and Luis Tapia. RJO thanks Daniel Nagasawa and Lucas Turner for their support during the McDonald observing runs, Dr. James Long for the useful discussion on the proper interpretation of a null hypothesis and Don Carona for his help at the Texas A\&M Campus Observatory. KMS was supported by NSF grant AST-1263034, ``REU Site: Astronomical Research and Instrumentation at Texas A\&M University." This work makes use of observations from the LCOGT network. Finally, the authors would like to thank the LCOGT TAC for the generous amount of telescope time provided.

\bibliographystyle{apj}
\bibliography{references}
\end{document}